\documentclass[%
 reprint,
 floatfix,
 aps,prl,
 showpacs,
 superscriptaddress,
]{revtex4-1}

\usepackage[usenames,dvipsnames]{xcolor}
\usepackage{graphicx}
\usepackage{dcolumn}
\usepackage{bm}
\usepackage{amsmath}
\usepackage{amssymb}
\usepackage{multirow}
\usepackage{hyperref}

\begin{document}


\title{Schottky barrier formation and band bending revealed by first principles calculations} 

\author{Yang Jiao}
\affiliation{Department of Applied Physics, Chalmers University of Technology, G\"oteborg, SE-412 96, Sweden.}
\author{Anders Hellman}
\email{anders.hellman@chalmers.se}
\affiliation{Department of Applied Physics, Chalmers University of Technology, G\"oteborg, SE-412 96, Sweden.}
\author{Yurui Fang}
\affiliation{Department of Applied Physics, Chalmers University of Technology, G\"oteborg, SE-412 96, Sweden.}
\author{Shiwu Gao}
\email{swgao@csrc.ac.cn}
\affiliation{Beijing Computational Science Research Center, Zhongguancun Software Park II, No. 10 Dongbeiwang West Road, Haidian District, Beijing, 100094, China.}
\author{Mikael K\"all}
\affiliation{Department of Applied Physics, Chalmers University of Technology, G\"oteborg, SE-412 96, Sweden.}


\date{\today}

\begin{abstract}

An atomistic insight into potential barrier formation and band bending at the interface between a metal and an n-type semiconductor is achieved by \textit{ab initio} simulations and model analysis of a prototype Schottky diode, {\it i.e.}, niobium doped rutile titania in contact with gold (Au/Nb:TiO$_2$). The local Schottky barrier height is found to vary between 0 and 1.26~eV depending on the position of the dopant. The band bending is caused by a dopant induced dipole field between the interface and the dopant site, whereas the pristine Au/TiO$_2$ interface does not show any band bending. These findings open the possibility for atomic scale optimization of the Schottky barrier and light harvesting in metal-semiconductor nanostructures. 

\end{abstract}


\maketitle

The presence of a 
Schottky barrier (SB) between a semiconductor and a metal is of paramount importance to  numerous application fields, including electronics~\cite{Ozpineci-2011}, photovoltaics~\cite{McFarland-2003,Nishijima-2010}, photocatalysis~\cite{Tian-2005,Clavero-2014,Nishijima-2012} and gas sensors~\cite{Gergen-2001,Park-2008,Lee-2015}. Schottky barrier physics has been a subject of intense investigation for several decades, but has recently received renewed substantial attention in two areas: i) the emergence of novel Schottky devices in plasmonics for photocurrent generation, photo detection and solar light harvesting~\cite{Clavero-2014,Brongersma-2015,Mubeen-2013}; and ii) the development of quantum-scale metal-semiconductor structures, pushed by the ever present need to further minimize and optimize electronic devices~\cite{Ifflander-2015,Padilha-2015,Durcan-2015,Lee-2014,Suyatin-2014,Gammon-2013}. Continued development of these areas could be greatly facilitated by an atomistic understanding of SB-based processes.

The quantum transmission of electrons or holes across the SB is determined by two quantities: the barrier height and, more importantly, the decay length of the band bending. Together, these quantities determine the probability of transmission and the energy distribution of hot carriers across the metal-semiconductor interface.  
The conventional SB model assumes a uniform charge depletion region on the semiconductor side and a charge accumulation layer localized at the interface~\cite{Sze-1985,Rhoderick-1988}, resulting in a parabolic bending of the semiconductor bands. The decay length of the band bending has been believed to be on the order of 10 nanometers for typical dopant concentration ($10^{19}~\text{cm}^{-3}$). 
However, recent advances in nanotechnology~\cite{Ifflander-2015,Kaiser-1988,Bell-1988}, which has made it possible to control and characterize the SB at the nanometer scale, have revealed important deviations from predictions made from the homogeneous Schottky barrier height (SBH) model~\cite{Durcan-2015}. Instead, the results, which depend on materials properties, dopant compositions and concentrations, have been qualitatively interpreted in the inhomogeneous SBH model~\cite{Tung-1992,Tung-2001}, although the current lack of a complete atomistic picture severely limits the possibility to achieve a quantitative understanding of SB formation. 
In particular, conventional models all assume a uniform dopant distribution, and it remains unclear how the composition and atomic structure of the semiconductor affect the electronic structure, i.e., barrier height and band bending, on the atomic level.  

Here, we report first-principles calculations of a prototype Schottky diode, {\it i.e.}, the Au/TiO$_2$ interface. A substitutional niobium dopant ($Nb_{Ti}$) was introduced to model discrete defects in the n-type semiconductor. The concentration and spatial distribution of the $Nb_{Ti}$ can be controlled experimentally~\cite{Morris-2000,Atanacio-2014,Sheppard-2012,Sheppard-2013} and Nb dopants are known to induce small lattice relaxations~\cite{Morgan-2009}, which makes this particular system suitable for atomistic simulations. We found that the pristine Au/TiO$_2$ interface has a relatively large barrier height, but shows no band bending. The band bending and decay length is instead determined by the precise locations of the dopant. Our results also show that the band bending is inhomogeneous and highly localized to the defect region. We calculated the dopant position dependent barrier height and show that it can be qualitatively understood by the deep level (DL) barrier model~\cite{Tung-2014}.
Our results reveal the origin and nature of inhomogeneity of the SBH and shed light on the mechanisms of electron transmission across the metal-semiconductor interfaces. 

All the calculation in this work was done using the Vienna Ab initio Simulation Packages
(VASP)~\cite{Kresse-1993,Kresse-1994,Kresse-1996-cms,Kresse-1996}
with the projector augmented wave (PAW) method~\cite{Blochl-1994,Kresse-1999} 
and the PBE~\cite{Perdew-1996,Perdew-1997} exchange-correlation functional 
in the generalized gradient approximation (GGA). The Coulomb correlation of the Ti $3d$ orbitals were treated in the GGA+U scheme with an effective on-site Coulomb repulsion U=10~eV, which was calibrated to the band gap and energetic position of the defect states. 
This U parameter results in a 3.28~eV indirect band gap at $M-\Gamma$ and a 3.36~eV
direct band gap at $\Gamma$ for bulk rutile TiO$_2$. The Nb defect states are located below the conduction band minimum (CBM).  (see Supplemental Material~\cite{SI} for calculation details and the choice of U parameters). 

\begin{figure}
\includegraphics[scale=1.0]{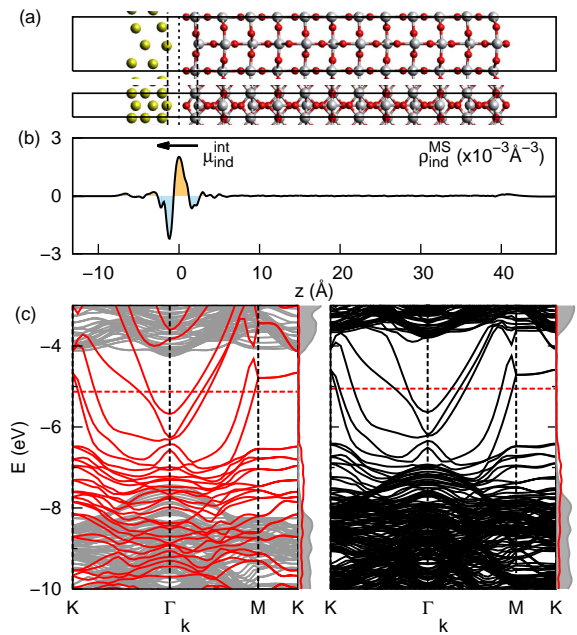}
\caption{\label{fig:cry}
The geometry and electronic structure of the pristine interface show that the TiO$_2$ band shifts up owing to the influence of the induced interface dipole.
(a) Schematic of the Au (112) and TiO$_2$ rutile (110) interface used in the calculations.  The boxes show the cell employed in the periodic boundary calculations.
(b) The plane (parallel to the slab) averaged induced charge density. Red and blue indicate charge accumulation and depletion.
The induced charge density gives rise to a dipole at the interface.
(c) The band structure of isolated (left) and combined (right) Au and rutile TiO$_2$ slab.
The red dashed line is the Fermi level. The side columns show the density of states (DOS) of the Au and TiO$_2$ components.
 The TiO$_2$ bands shift up by 0.4~eV owing to the contact with Au.
}
\end{figure}

In order to explore the effect of chemical composition on the SB, a large number of different atomic
structures were calculated based on the lattice alignment and 
orientation of Au nanoparticles on thin film rutile (110) TiO$_2$~\cite{Cosandey-2013}. 
In epitaxial growth, the Au[110] close-packed direction 
is always found to be parallel to the TiO$_2$[001] direction with a lattice mismatch 
as small as 0.4\%. Different epitaxial layers are obtained by rotating around the [110] axis. In the following, we focus on the Au(112)/TiO$_2$(110) interface, which has been observed 
after high temperature deposition \cite{Cosandey-2001}. In this case, 
one Au atom is located on top of $\mathrm{Ti}_{5c}$ in the
rutile $\mathrm{TiO}_2$(110) surface (Fig.~\ref{fig:cry}(a))~\cite{Cosandey-2001}.
We modeled the interface in a slab geometry using a 12 O-Ti-O trilayers (40~\AA~ thick) in contact with 3 layers (5~\AA~thick) of Au(112). The periodic slabs were separated by a 15~\AA~ thick vacuum region (Fig.~\ref{fig:cry}(a)), and dipole correction was used to alleviate the image interactions.
The work function of the Au slab ($\phi_m$) and
the electron affinity of the TiO$_2$ slab ($\chi_s$) were calculated to be  
 5.13~eV and 4.27~eV, 
which is in good agreement with experimental values (5.2~eV~\cite{Butler-1978} and 4.3~eV~\cite{Hansson-1978}, 
respectively). The Nb-dopant was introduced by replacing one of the Ti atoms in the lattice ($Nb_{Ti}$). In addition, the effect of changing the concentration of the Nb-dopants was checked by increasing the surface
unitcell to $(1\times2)$ and $(2\times3)$ using a 5-layer TiO$_2$ slab.

The electronic structure of the pristine interface is shown in Fig.~\ref{fig:cry}.
The lower panels show the band structures of the isolated
Au and TiO$_2$ slabs (left) and the interface formed upon contact (right). 
The interface band structure reveals a rigid upward shift of about 0.5~eV for both the valence and conduction bands of TiO$_2$, 
while the Fermi level and Au bands change by less than 0.1~eV. 
This difference is caused by charge polarization, as shown in panel (b). 
We analyzed this effect using the contact induced charge density
$\rho_{ind}^{MS}$ and doping induced charge density $\rho_{ind}^{D}$ defined as:
\begin{align}
 \rho_{ind}^{MS} & =\rho_{MS}-(\rho_{M}+\rho_{S}), \\
 \rho_{ind}^{D}  & =\rho_{D-MS}-(\rho_{MS}-\rho_{Ti}+\rho_{Nb}), \\
 \mu_{ind}^{int} & = \int^{int} \rho_{ind}^{MS/D} (z-z_0) dz. \label{eq:dind} 
\end{align}
Here, $\rho_{MS}$ and $\rho_{D-MS}$ are the charge densities of the undoped and doped Au/TiO$_2$ interfaces, respectively, 
$\rho_{M}$  and $\rho_{S}$ are those of the isolated metal and semiconductor slabs, while
$\rho_{Ti}$  and $\rho_{Nb}$ are the charge densities of the free atoms.
The areal density of the interface induced dipole $\mu_{ind}^{int}$ 
was calculated by integration over an interface range of
5~\AA~and found to be $\mu_{ind}^{int} = - 0.014 D\text{\AA}^{-2}$.
The negative sign means that the dipole points from TiO$_2$ to Au (Fig.~\ref{fig:cry}(b)).
The charge transfer between Au and undoped TiO$_2$ is found to be less than $0.05$~e/cell, as shown by a Bader analysis. The induced dipole results in a SBH of 1.26~eV, which is 0.4~eV 
larger than the difference between the work function and the electron affinity 
of the isolated Au and TiO$_2$ slabs.
The band gap of TiO$_2$ is 3.3~eV, that is the same as in the bulk calculation.
The interface position ($z=0$) was chosen at the electronic potential
maximum between the TiO$_2$ and Au layers. 
Importantly, the conduction band edge is found to be flat throughout the semiconductor slab (Fig. S3).

\begin{figure}
\includegraphics[scale=1.0]{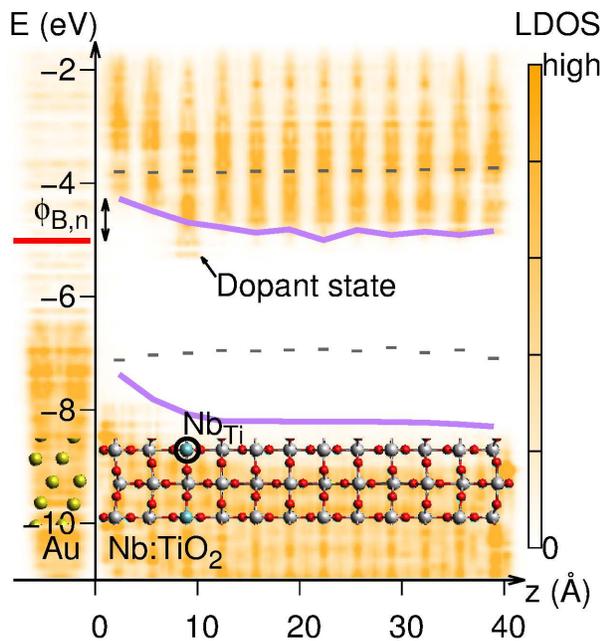}
\caption{\label{fig:abs}
Illustration of Schottky barrier formed at a Au/Nb:TiO$_2$ interface.
The color map shows the local density of states (LDOS), as obtained from the DFT+U calculations, along the direction normal to the interface. 
The Schottky barrier height for n-type doping ($\Phi_{B,n}$) is defined as the energy difference between the Fermi energy (red line) and the conduction band minimum (upper purple line) at the interface.
The Nb-dopant is marked as $Nb_{Ti}$ in the inserted structure plot.
The dopant state below the bottom of the conduction band traps most of the excess electron donated by $Nb_{Ti}$.
The band edges of the pristine TiO$_2$ are also shown (gray dashed lines).
The band bending is caused by the dopant induced charge polarization. 
 }
\end{figure}

The results above show that the pristine TiO$_2$/Au interface
does not exhibit any band bending in the semiconductor region,
and we therefore hypothesized that the atomic scale SBH
inhomogeneity and band bending instead are caused by the dopant.
Figure~\ref{fig:abs} shows a typical case with a Nb-dopant placed
two layers beneath the interface. Here the SBH ($\Phi_{B,n}$) is
defined as the energy difference between the Fermi level ($E_F$)
and the conduction band minimum (CBM) of the TiO$_2$ layer at the
interface. We found that both the potential profile and the SBH
critically depend on the position of the dopant. When the
$Nb_{Ti}$ is located at the contacting layer, the defect state
density is high enough to pin the Fermi level, which is very
close to the bottom of the conduction band, as found in previous
studies~\cite{Marri-2008}. In this case, the SBH is significantly
reduced, and the band bending is the same as for the pristine
case, i.e. it is essentially flat. However, as the dopant is
moved away from the interface, the dopant induced charge is
distributed nonuniformly between the interface and the dopant,
leading to a lowering of the potential towards the dopant, i.e.,
a clear band bending. This trend starts from the second layer
(dopant-interface distance $d_D=5.9$~\AA) and becomes more
prominent as the dopant moves to the other end of the slab at
12\textit{th} layer ($d_D=39.2$~\AA). Thus, our results clearly
show that the local dopant confine the SBH and dictate the band
bending.

\begin{figure}
\includegraphics[scale=1.0]{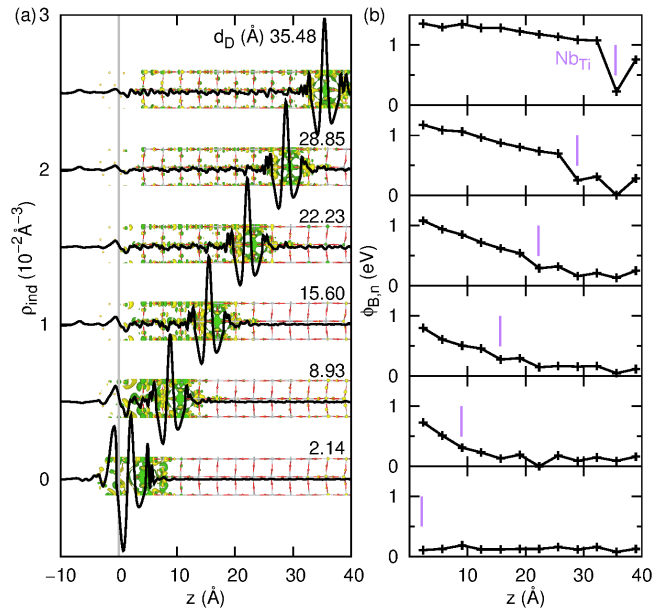}
\caption{\label{fig:indchg}
The dopant-interface distance ($d_D$) dependence of the induced charge and band bending.
(a) $Nb_{Ti}$ induced charge density. 
The curves are the plane averaged value and the color inserts are the isosurface with isovalue  $=10^{-3}~\text{\AA}^{-3}$. Yellow (+), Green (-).
The induced interface dipole is inversely proportional to $d_D$.
(b) Band bending in TiO$_2$. 
The points are the local conduction band minimum (CBM) in each layer in relation to the Fermi energy. 
The vertical bar indicates the position of Nb-dopant in each calculation. 
The TiO$_2$ band shifts up while approaching the interface.
}
\end{figure}

The variation in charge and potential with dopant-interface
distance are analyzed in Fig.~\ref{fig:indchg}. With the
Nb-dopant positioned between the second layer and the fifth
layer, the band bending is nearly parabolic. This is consistent
with the uniform dopant SB model. However, it becomes essentially
linear if the Nb-dopant is located beyond the fifth layer, i.e.
$d_D>16~\text{\AA}$. The magnitude of the induced interface
dipole is found to be inversely proportional to the
dopant-interface distance,
$\mu_{ind}^{int}=0.066~\text{D\AA}^{-1}/(d_D-1.08~\text{\AA})$
(Fig. S4), and the dipole points from Au to Nb:TiO$_2$, while the
induced dipole around the dopant points in the opposite
direction. The charge transfer between the metal and the
semiconductor slabs was found to be negligible for the pristine
interface as well as for the doped cases (0.07~e/cell in case of
interfacial $Nb_{Ti}$, and less than 0.02~e/cell at other doping
positions). The excess electron contributed by the Nb-dopant is
instead mainly distributed on the Nb atom and the neighboring Ti
atoms along the [001] direction. This localized charge
distribution is consistent with previous DFT+U
calculations~\cite{Morgan-2009} and STM
experiments~\cite{Setvin-2014}. Further Fig.~\ref{fig:indchg}(b)
shows that the CBM of the layer containing the dopant was pinned
to the energy $E_F+\zeta$~eV, where $\zeta$ is in the range of
$0.2\sim0.5$~eV and varies with $d_D$. The same shift was
observed for the Ti 4s semi-core states (Fig. S5). Thus, we can
conclude that dopant induced charge polarization give rise to a
dipole field that is mainly responsible for the band bending.
We can further conclude that the SBH and band bending are highly inhomogeneous
(locally determined) and strongly dependent on the
dopant-interface distance within a range of a few nanometers.

\begin{figure}
\includegraphics[scale=1.0]{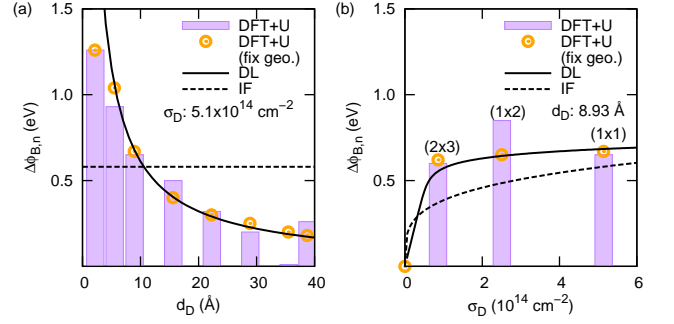}
\caption{\label{fig:dl}
Schottky barrier height reduction ($\Delta\Phi_{B,n}$) as a function of (a) dopant-interface distance ($d_{D}$) and (b) dopant areal concentration ($\sigma_{D}$). The histogram show the DFT+U results obtained with full geometry optimization. The circles are calculations with atomic coordinates fixed at the pristine interface geometry. The solid lines are calculated using the DL model (Eq.~\ref{eq:dl}) while the dashed lines are obtained from the uniform dopant Schottky model including an image force (IF) correction (Eq.~\ref{eq:if}). }
\end{figure}

The atomic picture described above is in sharp contrast to the conventional SB picture, which does not include any 
 dopant position dependence as it
 assumes that the band bending is caused by the electrostatic potential generated by uniform ionized defects in the depletion layer~\cite{Sze-1985,Rhoderick-1988}.  
The conventional potential profile consists of a parabolic term plus an image charge correction term~\cite{Sze-1985,Sze-1964}.  
The image force (IF) decreases the SBH by~\cite{Rhoderick-1988}:
\begin{equation}
\Delta \Phi_{B,n}^{IF} = \left[ \frac{e^6N_D}{8\pi^2\varepsilon_0^3\varepsilon_S^3} \left(\Phi_{B,n}^0-\zeta-k_BT\right)  \right]^{\frac{1}{4}}, 
\label{eq:if}
\end{equation}
where $e$ is the charge of the electron, $N_D$ is the dopant concentration, $\varepsilon_0$ and $\varepsilon_S$ are the vacuum permittivity and relative dielectric constant of semiconductor, respectively, $\Phi_{B,n}^0$ is the SBH in the absence of image charge correction, $\zeta$ is the energy difference between the CBM and the Fermi level, $k_B$ is Boltzmann's constant and $T$ is the temperature.  Using the geometry from our first principles calculations, i.e. one
$Nb_{Ti}$ in the long $(1\times1\times12)$ cell, we have $N_D=1.3\times10^{21}~\text{cm}^{-3}$. The depletion layer width is found to be 10~\AA~(25~\AA) with $\varepsilon_S=10$ ($\varepsilon_S=60$). The SBH reduction is then 0.58~eV for $\varepsilon_S=10$ (0.15~eV for $\varepsilon_S=60$) (Fig. S6), which deviates from the calculated SBH reduction (Fig.~\ref{fig:dl}), and does not explain the dependence on the dopant position (Fig.~\ref{fig:indchg}).  
This qualitative and quantitative discrepancy between the SBH prediction obtained from the uniform dopant Schottky model and our results clearly indicate the importance of an atomistic description of the interface. 

We now compare the DFT+U results with the 
alternative deep level (DL) model~\cite{Tung-2014}, which was developed to specifically incorporate the local barrier profile near the interface. The DL model assumes point charge donors with energy $(E_{DL})$  below the CBM. The SBH reduction depends on the charge donor-interface distance ($d_{DL}$) and the areal dopant density ($\sigma_{DL}$) according to~\cite{Tung-2014} 
\begin{equation}
\Delta\Phi_{B,n}^{DL} = \frac{e^2a\sigma_{DL}(1-f_{DL})}{\varepsilon_0 \varepsilon_S},  \label{eq:dl}
\end{equation}
where \\ $f_{DL}=\left\{ 1+\exp\left[  \frac{\Phi_{B,n}^0-e^2d_{DL}\sigma_{DL}(1-f_{DL})/(\varepsilon_0\varepsilon_S)-E_{DL} }{k_BT} \right] \right\}^{-1}$  is the Fermi-Dirac distribution function of the DL states. We extracted the relevant parameters from the DFT+U calculations (Fig.~\ref{fig:indchg}):  
$\Phi_{B,n}^0=1.26~\text{eV}$ is the SBH without DL states, $a=5$~\AA~is the characteristic width of the interface, 
$E_{DL}=0.2~\text{eV}$ is the DL energy below the CBM,
and we set $\varepsilon_S=10$ and $T=1000~K$.
The resulting SBH obtained from the DL model is plotted in Fig.~\ref{fig:dl} and is found to be in good agreement with our first principles calculations, except for the regions closest to the boundaries. To pin down the origin of this discrepancy, we also performed calculations for a case where the atomic structure of the Nb-doped slab were fixed at the pristine interface structure position (DFT+U (fix geo.) in Fig.~\ref{fig:dl}). 
The DL model then captures the main dopant dependent SBH feature extremely well, the only exception is the case when the dopant is located at the interfacial layer where the DL model is not applicable.

In conclusion, first principles calculations of Au/TiO$_2$
interfaces show that the SBH is tuned by interface dipoles induced by contact and dopant. The local barrier profile, i.e. the band bending, shift almost linearly
between the interface and the dopant layer. The barrier width is a-few-layer thick and depends on the dopant position. The reported experimental SBH for Au/TiO$_2$ interfaces lies in the range 0.9$\sim$1.2~eV~\cite{Tang-2003,Park-2008,Lee-2012}.   
Given that these measurements are macroscopically averaged, our
calculations are in very good agreement with experiment. 
In contrast to the uniform dopant Schottky model, the DL model is able to
account for the SBH reduction variation with dopant position. 
The conclusion and overall picture emerging from the present study should be generally applicable and highly relevant also to other metal-semiconductor systems. As such, they can serve as a basis and reference to further studies of internal electron emission and hot-carrier transport across metal semiconductor interfaces. In light of the rapid development of layer-controlled molecular beam epitaxy~\cite{Son-2010}, the results open up the possibility for atomic scale engineering and optimization of novel SB-based devices.

\begin{acknowledgments}
This research was supported by the Knut and Alice Wallenberg Foundation.
The computations were performed on resources provided by the Swedish
National Infrastructure for Computing (SNIC) at NSC.  
\end{acknowledgments}

\bibliographystyle{apsrev4-1}
\bibliography{sb}

\end{document}


\preprint{APS/123-QED}

\title{Supplementary information: Schottky barrier formation and band bending revealed by first principles calculations} 

\author{Yang Jiao}
\affiliation{Department of Applied Physics, Chalmers University of Technology, G\"oteborg, SE-412 96, Sweden.}
\author{Anders Hellman}
\email{anders.hellman@chalmers.se}
\affiliation{Department of Applied Physics, Chalmers University of Technology, G\"oteborg, SE-412 96, Sweden.}
\author{Yurui Fang}
\affiliation{Department of Applied Physics, Chalmers University of Technology, G\"oteborg, SE-412 96, Sweden.}
\author{Shiwu Gao}
\email{swgao@csrc.ac.cn}
\affiliation{Beijing Computational Science Research Center, Zhongguancun Software Park II, No. 10 Dongbeiwang West Road, Haidian District, Beijing, 100094, China.}
\author{Mikael K\"all}
\affiliation{Department of Applied Physics, Chalmers University of Technology, G\"oteborg, SE-412 96, Sweden.}


\date{\today}

\maketitle

\renewcommand{\theequation}{S\arabic{equation}}
\renewcommand{\thefigure}{S\arabic{figure}}
\renewcommand{\thetable}{S\arabic{table}}

\section{Computation method}

In the density functional theory (DFT) calculations,  
the plane wave cut-off energy was set to 400~eV. A Monkhorst-Pack
k-point mesh of $5\times11\times1$ centered at the $\Gamma$ point was
used for the Au/TiO$_2$ rutile (110) interface supercell. A more
condensed k-sampling ($9\times21\times1$) was adopted when
extracting the density of states (DOS).
The DFT+U calculations were performed using Dudarev’s scheme~\cite{Dudarev-1998}. The geometry optimizations were done on the GGA level. All of the atoms were relaxed in the geometry optimization processes, and were considered to be relaxed when the maximum forces on each atom was smaller than 0.01~eV/\AA. The calculated lattice constants of TiO$_2$ (rutile) and Au (fcc) are $a=4.64$~\AA, $c=2.96$~\AA, $u=0.305$
and $a=4.17$~\AA, respectively, in good agreement with experimental values. 

\section{U determination}

In the DFT+U calculations, the on-site Coulomb repulsion U parameter was carefully
calibrated by calculating the bandgap of
pristine rutile TiO$_2$ and comparing with results from both
G$_0$W$_0$ calculations and experimental measurements. The calculations were
performed using a cell with 6 atoms and a $9\times9\times11$
Monkhorst-Pack k-point sampling. The parameter was determined to be
$U(Ti3d)=10$~eV (Table ~\ref{tab:UTi}). Furthermore, with this
value the electron affinity of 12-layer TiO$_2$ rutile slab was
found to be 4.27~eV, which also agrees well with the experimental
value of 4.33~eV \cite{Butler-1978}.

\begin{table}[h]
\centering
\caption{The band gap ($E_{gap}$) in bulk rutile TiO$_2$ and electron affinity ($\chi_s$) of the 12-layer rutile (110) slab. (units in eV)}
\label{tab:UTi}
\begin{tabular}{p{2.5cm}p{1cm}p{1.5cm}p{1.5cm}p{1.5cm}}
\hline \hline
 \multirow{2}{*}{} & & $E_{gap}^{direct}$ & $E_{gap}^{indirect}$ & $\chi_s$ \\
                       & & ($\Gamma\to\Gamma$) & ($M\to\Gamma$) &   \\
\hline
 \multirow{6}{*}{\makecell{DFT+U\\ $U(Ti3d)=$ }}& 0  & 1.88  & - & \\
 &  2  & 2.12  & -     &  \\
 &  4  & 2.39  & -     &  \\
 &  6  & 2.69  & 2.67  &  \\ 
 &  8  & 2.99  & 2.95  &  \\
 & 10  & 3.31  & 3.25  & 4.27  \\
$G_0W_0$ &   & 3.41 & 3.34 &  \\
Exp. &   &  \multicolumn{2}{c}{3.3~\cite{Tezuka-1994}} &  4.33~\cite{Butler-1978} \\
\hline
\end{tabular}
\end{table}

\begin{figure}
\centering
\includegraphics[width=3.375in]{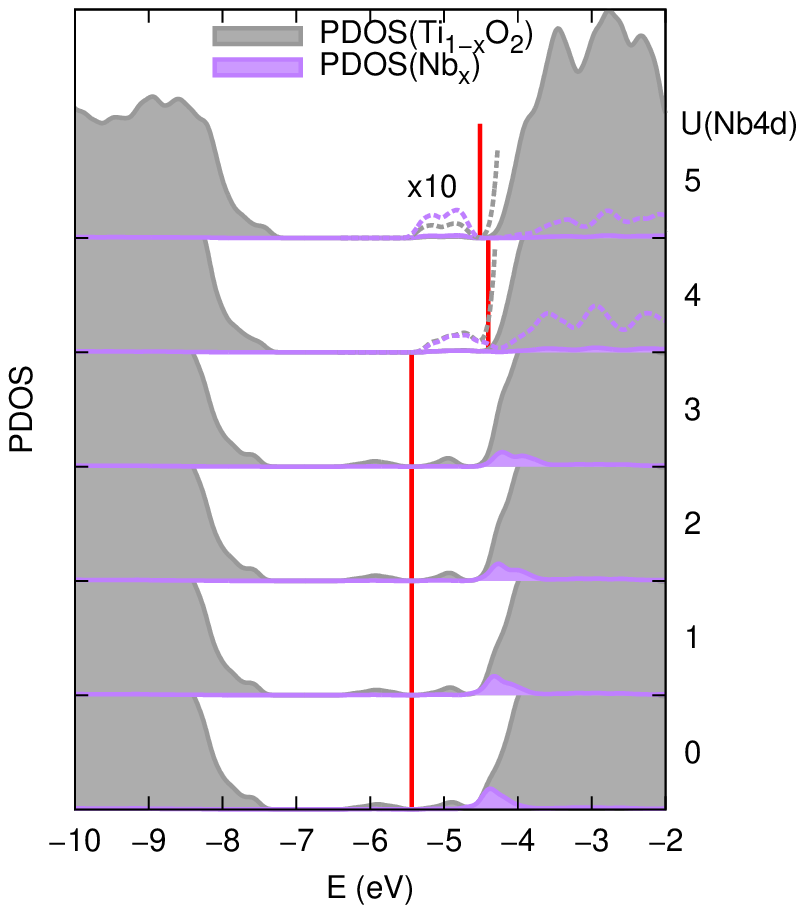}
\caption{The projected density of states (PDOS) of $Nb_xTi_{1-x}O_2$ on host ($Ti_{1-x}O_2$) and dopant ($Nb$). 
The $U(Ti3d)$ parameter was fixed at 10~eV and the $U(Nb4d)$ varied from 0 to 5 eV. 
The energy was aligned according to Ti 4s semi core orbital energy. The vertical red bars mark the Fermi energy in each calculations. 
}
\label{fig:nbpdos}
\end{figure}

\begin{figure}
\centering
\includegraphics[width=3.375in]{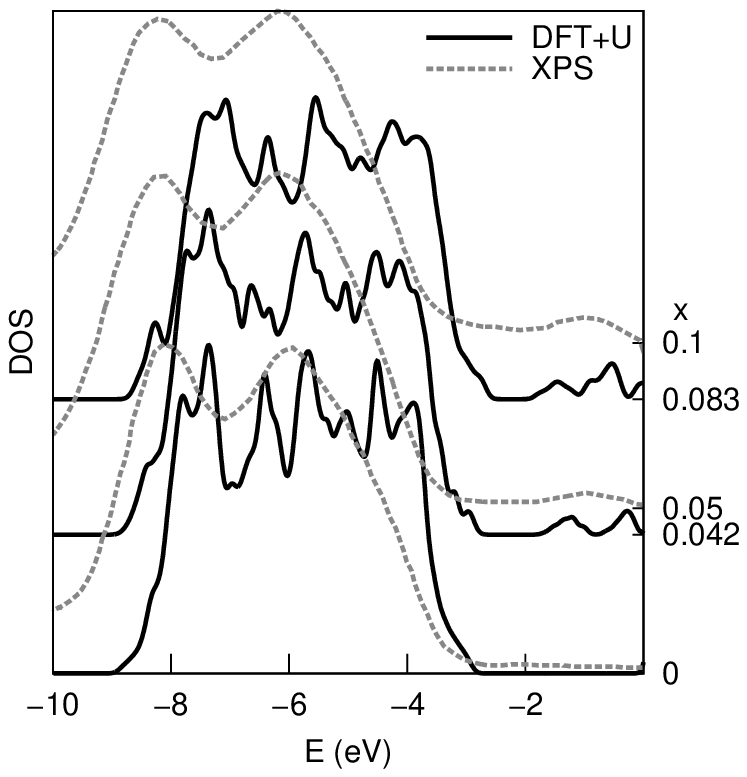}
\caption{The calculated density of states (DOS) using $U(Ti3d)=10$~eV and $U(Nb4d)=0$~eV compared with XPS experiments~\cite{Morris-2000}. The energy was aligned to the conduction band minimum.}
\label{fig:nbdos}
\end{figure}

The $U(Nb4d)$ was then determined by keeping the $U(Ti3d)$ fixed at 10~eV
and calculating the dopant state with $U(Nb4d)$ varied from 0~ 
eV to 5 eV (Table \ref{tab:UNb}). A supercell of
$2\times3\times2$ along [1$\bar{1}$0], [001] and [110]
directions was considered. The Nb dopant occupies one of the host
Ti sites within the supercell~\cite{Morris-2000}, thus forms
$\text{Ti}_{1-x}\text{Nb}_x\text{O}_2$ with $x=0.042$, which is
equivalent to our 12-layer TiO$_2$ calculations (a dopant concentration of
$1.3\times10^{21}~\text{cm}^{-3}$ and 4.7~wt\%).
Higher concentration of Nb-dopant was also calculated using the same supercell but with two
Ti atoms replaced by Nb. With $U(Nb4d)\le 3$, the charge carrier
distribution (Table~\ref{tab:UNb}) and states at the bottom of
conduction band (Fig.~\ref{fig:nbpdos}) were not sensitive to the
 varied value of the $U(Nb4d)$ parameter. 
When $U(Nb4d)$ increases from 0 to 3~eV, a dopant state locates 
at $1.3\sim1.5$~eV below the conduction band edge and about 0.5~eV below the Fermi level, while the extra charge was trapped on neighbor Ti atoms along [001] direction.  
The DOS below the conduction
band edge (CBE) was found compatible with x-ray photoemission
spectra (XPS)~\cite{Morris-2000} (Fig.~\ref{fig:nbdos}), and the
charge transfer to the neighboring Ti atoms along [001] direction
created a polaron in agreement with previous
studies~\cite{Morgan-2009}. For U(Nb4d)$\geq$ 4~eV, the dopant
state is upshift to just below the conduction band edge, which
results in remarkable changes in the charge distribution (Table
\ref{tab:UNb}). Based on the consideration of the defect state 
energy distribution
and charge carrier localization, $U(Nb4d)=0$ was sufficient
in this work.

\begin{table}[h]
\centering
\caption{The space distribution of the extra electron introduced by $Nb_{Ti}$ on Nb atom and the two nearest neighbor Ti atoms along [001] directions. $\times 2$ indicates the charge distributions on the two nearest neighbor Ti atoms are equal.
 }
\label{tab:UNb}
\begin{tabular}{p{2cm}p{2cm}p{2cm}}
\hline \hline
 U(Nb4d)    &  $n_{Nb}$   & $n_{Ti}$  \\
\hline
 0 &  0.101  & $0.363\times 2$    \\
 1 &  0.091  & $0.368\times 2$    \\
 2 &  0.082  & $0.374\times 2$   \\
 3 &  0.073  & $0.378\times 2$    \\
 4 &  0.451  & $0.154\times 2$    \\
 5 &  0.622  & $0.064\times 2$   \\
\hline
\end{tabular}
\end{table}

\section{Step-like barrier at pristine interface}

\begin{figure}[H]
\includegraphics[width=3.375in]{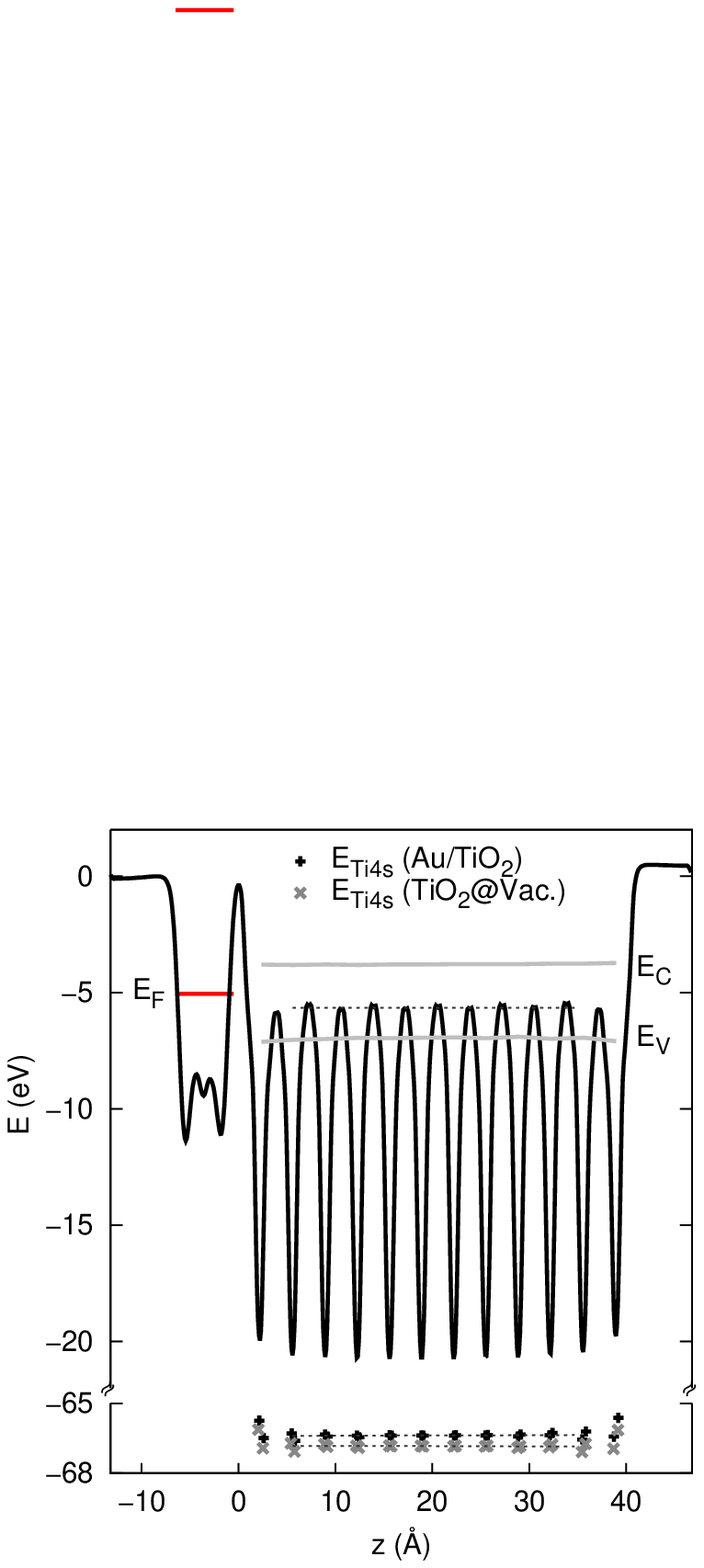}
\caption{The step-like barrier at the pristine Au/TiO$_2$ interface. 
The solid curve shows the plane averaged electrostatic potential of the Au/TiO$_2$ slab. The gray lines are the connection of the band edge of each TiO$_2$ layer.
The dots are the energy level of Ti 4s electrons in free standing and Au attached TiO$_2$ slabs. 
The interface to the Au slab upshifts the energy level of TiO$_2$ slab by 0.4~eV as a result of the interface electron polarization.
}
\label{fig:pot}
\end{figure}

Pristine Au/TiO$_2$ interface forms step-like barrier 
 as shown in Fig.~\ref{fig:pot}. The electron
polarization at the interface forms a plane of induced dipole
(Fig. 1(b)). The potential in the semiconductor side 
upshifts by a constant value. Figure~\ref{fig:pot} shows the
electrostatic potential, band edge of each TiO$_2$ layer and the
energy level of Ti 4s semi core orbital. The dotted lines are guide
lines to show the potential flatness in the interior of TiO$_2$
slab.

\section{Nb-dopant induced dipole and band bending}

\begin{figure}[H]
\includegraphics[width=3.375in]{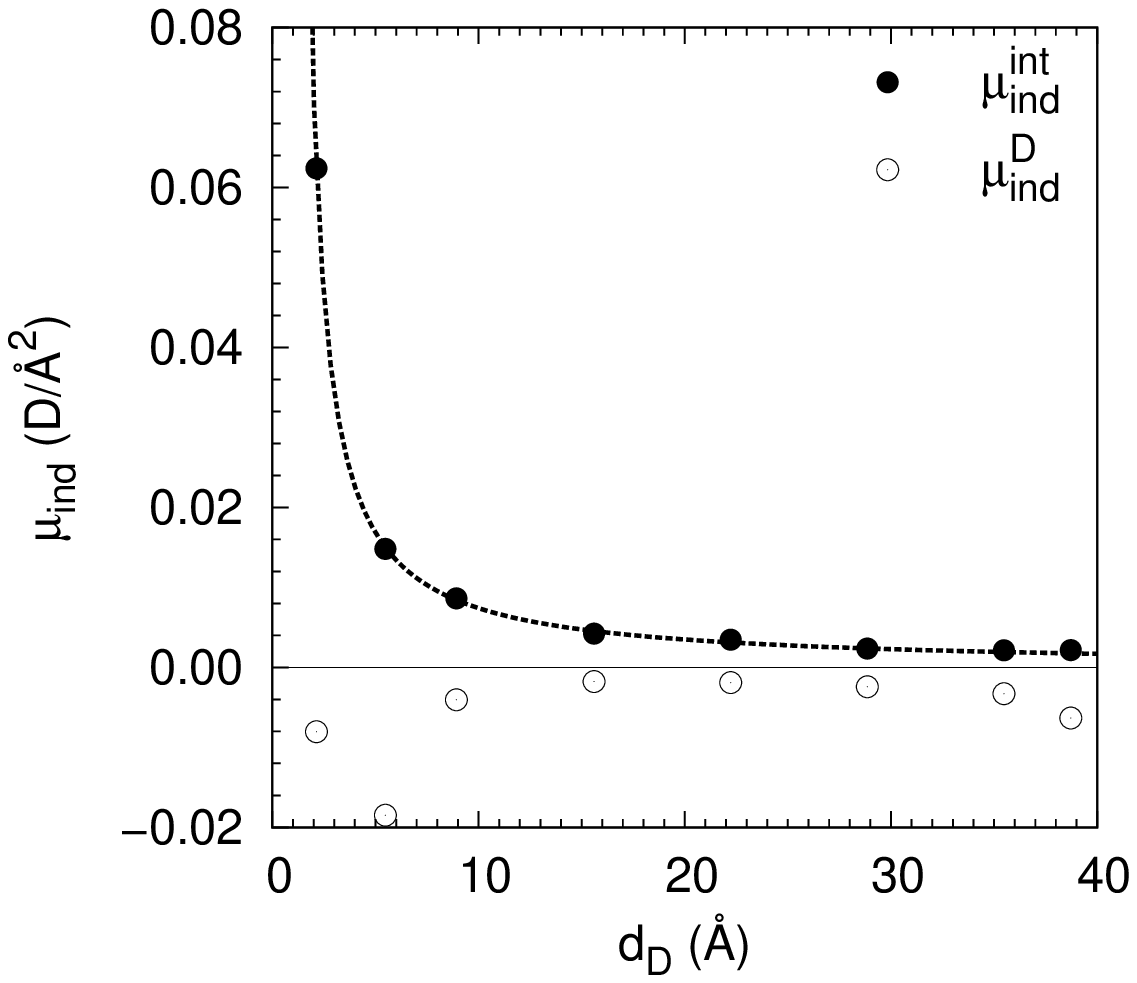}
\caption{The induced dipole at the interface ($\mu_{ind}^{int}$) and around the Nb dopant ($\mu_{ind}^{D}$) as a function of dopant interface distance ($d_D$). The dotted curve is fitted by $\mu_{ind}^{int}=0.066~\text{D\AA}^{-1}/(d_D-1.08~\text{\AA})$.}
\label{fig:dip}
\end{figure}

The Nb-dopant induces a dipole field between the interface and the
dopant layer. The magnitude was calculated by integrating of Eq.
3 over the interface range and dopant layer. The induced dipole
at the interface ($\mu_{ind}^{int}$) was found to be in reverse proportional to the
dopant interface distance ($d_D$) as shown in Fig.~\ref{fig:dip}.

\begin{figure}[H]
\includegraphics[width=3.375in]{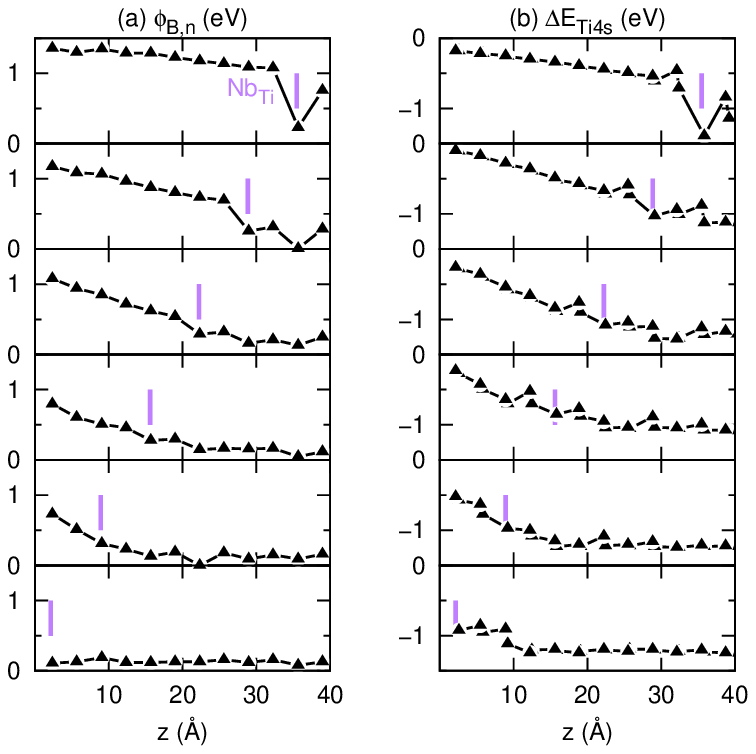}
\caption{The band bending in TiO$_2$ varied with $Nb$ dopant interface distance ($d_D$).
(a) The layer localized conduction band minimum (CBM) 
referred to the Fermi energy. (b) The dopant induced energy shift of the Ti 4s semi core state. The vertical bars indicate the position of Nb-dopant in each calculation. Both curves show that the TiO$_2$ bands shift up while approaching the interface.}
\label{fig:bb}
\end{figure}

The band bending is also presented by the shift of Ti 4s semi core orbital energy. 
They show the same trend with the semiconductor band edges (Fig. ~\ref{fig:bb}).

\newpage
\section{Schottky barrier height predicted by models}

\begin{figure}[H]
\includegraphics[width=3.375in]{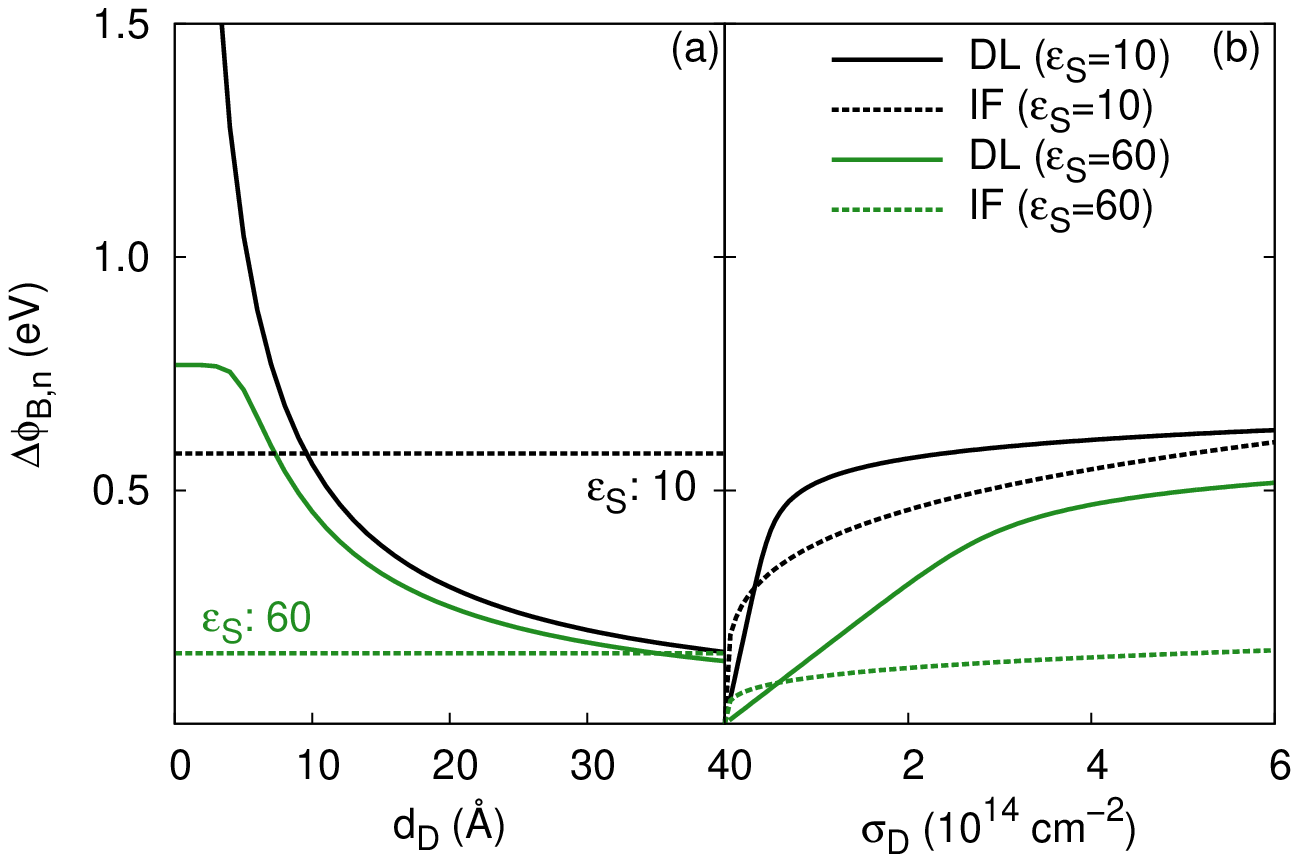}
\caption{
The decrease of the SBH as a function of the dopant interface distance (a) and dopant areal concentration (b) using model simulations with low and high dielectric constants ($\varepsilon_S$).
}
\label{fig:dl}
\end{figure}

In the analysis of SBH with image force (IF) correction and deep level (DL) model, 
 the dielectric constant $\varepsilon_S$ was treated as a parameter. 
DL model with $\varepsilon_S=10$ shows good agreement with our first principles calculations (Fig. 4).  As TiO$_2$ has dielectric constant of $\varepsilon_S^{1000K} = 97$ and $58$ in the c and a directions~\cite{Parker-1961}, we also calculated the model predicted SBH reduction $\Delta\Phi_{B,n}$ with dielectric constant of $\varepsilon_S=60$ as shown in Fig.~\ref{fig:dl}.

The uniform dopant Schottky model with IF correction (Eq. 4) is more sensitive to the
dielectric constant as compared to localized DL model. With $\varepsilon_S=60$
the SBH reduction is always less than $0.2$~eV in the considered
density range. And the depletion layer width expands from 10~\AA ($\varepsilon_S=10$) to 25~\AA ($\varepsilon_S=60$) under the dopant density of $N_D=1.3\times10^{21}\text{~cm}^{-3}$ . 

\bibliographystyle{apsrev4-1}
\bibliography{sb}